\documentstyle[aps,epsfig]{revtex}
\begin{document}
\title{
Nuclear  absorption  and  anomalous $J/\psi$  suppression in Pb+Pb
collisions}

\author{\bf A. K. Chaudhuri\cite{byline}}
\address{ Variable Energy Cyclotron Centre\\
1/AF,Bidhan Nagar, Calcutta - 700 064\\}
\date{\today}

\maketitle

\begin{abstract}

We  have  studied  the  $J/\psi$  suppression  in 158 GeV/c Pb+Pb 
collisions at CERN SPS. $J/\psi$ production is assumed  to  be  a
two  step  process,  (i)  formation  of $c\bar{c}$ pair, which is
accurately calculable in QCD and (ii) formation of $J/\psi$ meson
from   the   $c\bar{c}$   pair,   which   can   be   conveniently
parameterized.  In a pA/AA collision, as the $c\bar{c}$ pair pass
through the nuclear medium, it gains relative square momentum. As
a result, some of the $c\bar{c}$ pairs can gain  enough  momentum
to  cross  the  threshold  to become open charm meson, leading to
suppression in pA/AA  collisions.  The  model  without  any  free
parameter   could   describe  the  of  NA50  data  on  centrality
dependence of the ratio's; $J/\psi$ over Drell-Yan, $J/\psi$ over
minimum bias and also the Drell-Yan over minimum bias. The  model
was  used to predict $J/\psi$ suppression at RHIC energy. At RHIC
energy, hard processes may  be  important.  With  hard  processes
included,  $J/\psi$'s  are strongly suppressed, in agreement with
other model calculations. We also show that centrality dependence
of $J/\psi$ over minimum bias ratio can be used to determine  the
fraction of hard processes in the collision.

\end{abstract}


\section{Introduction}

One  of the predictions of quantum chromodynamics is the possible
existence  of  a  deconfined  state  of  quarks  and  gluons.  In
relativistic  heavy  ion  collisions,  under  certain  conditions
(sufficiently  high  energy  density  and  temperature)  ordinary
hadronic  matter  (where  quarks  and  gluons  are  confined) can
undergo a phase transition to deconfined matter,  commonly  known
as  quark  gluon  plasma.  Over the years, nuclear physicists are
trying to produce and detect this new phase of matter,  first  at
CERN  SPS  and  now  at  Relativistic  Heavy Ion Collider (RHIC).
Indeed, it has even been claimed that evidence of  new  state  of
matter   is   already   seen   at   Pb+Pb   collisions   at  CERN
SPS\cite{he00}. The claim was based mainly  on  the  experimental
$J/\psi$   suppression   and   strangeness  enhancement.  Due  to
intrinsic ambiguities in disentangling  similar  effects  arising
from  hadronic  processes,  the  claim is considered to be rather
controversial.

$J/\psi$  suppression  is  recognized  as  an  important  tool to
identify the possible phase  transition  to  quark-gluon  plasma.
Because  of  the large mass of the charm quarks, $c\bar{c}$ pairs
are produced on a short time  scale.  Their  tight  binding  also
makes  them  immune  to final state interactions. Their evolution
probes the state of matter in the early stage of the  collisions.
Matsui  and  Satz  \cite{ma86}  predicted  that  in  presence  of
quark-gluon plasma  (QGP),  due  to  screening  of  color  force,
binding of $c\bar{c}$ pairs into $J/\psi$ meson will be hindered,
leading  to  the  so  called  $J/\psi$  suppression  in heavy ion
collisions  \cite{ma86}.  Over  the  years  several  groups  have
measured the $J/\psi$ yield in heavy ion collisions (for a review
of  the data and the interpretations see Refs. \cite{vo99,ge99}).
In brief, experimental data do  show  suppression.  However  this
could  be attributed to the conventional nuclear absorption, also
present in $pA$ collisions.

The latest data obtained by the NA50 collaboration \cite{na50} on
J/$\psi$ production in Pb+Pb collisions at 158 A GeV is the first
indication  of the anomalous mechanism of charmonium suppression,
which goes beyond  the  conventional  suppression  in  a  nuclear
environment.  The  ratio  of  $J/\psi$ yield to that of Drell-Yan
pairs decreases faster with $E_T$ in the most central  collisions
than  in  the  less  central ones. It has been suggested that the
resulting pattern can be understood in a  deconfinement  scenario
in  terms  of  successive  melting  of  charmonium  bound  states
\cite{na50}. Blaizot {\em et al.} \cite{bl00} have shown that the
data  can  be  understood  as  an  effect  of  transverse  energy
fluctuations  in  central  heavy  ion  collisions.  Introducing a
factor $\varepsilon=E_T/E_T(b)$ , assuming that  the  suppression
is  100\%  above  a threshold density (a parameter in the model),
and smearing the threshold density (at  the  expense  of  another
parameter)  the  best fit to the data was obtained. Extending the
Blaizot's  model  to  include  fluctuations  in  number   of   NN
collisions at a fixed impact parameter, NA50 data could be fitted
with  a  single parameter, the threshold density, above which all
the $J/\psi$ mesons melt \cite{ch01a,ch01b}. Assumption that  all
the  $J/\psi$  mesons  melt above a threshold density, implicitly
assume that a QGP like environment is produced in the collisions.
NA50 data could also be explained in the  conventional  approach,
without   invoking  QGP  like  scenario.  Capella  {\em  et  al.}
\cite{ca00} analyzed the data  in  the  comover  approach.  There
also,  the  comover  density  has  to  be  modified by the factor
$\varepsilon$. Introduction of this  adhoc  factor  $\varepsilon$
can be justified in a model based on excited nucleons represented
by strings \cite{hu00}. Recently we have shown that the NA50 data
is  well  explained  in  a  QCD  based  model \cite{qiu98}, where
formation of $J/\psi$ from $c\bar{c}$ pair is suppressed, due  to
gain  in  the  relative  square momentum of $c\bar{c}$ pair as it
moves  through  a  nuclear  medium  \cite{ch02}.  Some   of   the
$c\bar{c}$  pair can gain enough square momenta to cross the open
charm threshold, thereby reducing the $J/\psi$ cross section. The
model predicted large suppression of $J/\psi$ at RHIC energy,  in
agreement with other model predictions.

In  the  present  paper  we have extended our earlier analysis of
NA50  data  of  transverse  energy  dependence  of  $J/\psi$   to
Drell-Yan  ratio  \cite{ch02},  to  include the transverse energy
dependence of $J/\psi$ over MB ratio and the DY  over  MB  ratio.
Recently Capella et al \cite{ca01} observed that on the average, a
$J/\psi$  and  also  a  DY event has $E_T \sim$ 3 GeV less than a
minimum bias event. $E_T$ loss in a  $J/\psi$(DY)  event,  though
small,  can  have  larger  effect  beyond  the  knee of the $E_T$
distribution. We have also studied the effect of  $E_T$  loss  in
$J/\psi$  and in DY events. We confirm the analysis of Capella et
al \cite{ca01} that in the ratio $J/\psi$  over  DY,  $E_T$  loss
effect  gets cancelled. But the effect is manifested in the ratio
of $J/\psi$ over MB as well as in the ratio of DY over MB.  Next,
we  apply  the  model  to  predict  $E_T$  dependence of $J/\psi$
suppression in Au+Au collisions at RHIC energy. At  RHIC  energy,
hard scattering may be important. Suppression increases with hard
scattering.  It  is  also  shown  that $J/\psi$ over minimum bias
ratio can be used to obtain fraction of hard  scattering  in  the
collision.

The  plan  of  the  paper is as follows: in section 2, we briefly
present the model of $J/\psi$ production  and  absorption  in  AA
collisions.  In  section  3,  we  obtain  the  transverse  energy
distribution  in  Pb+Pb  collisions.  In  section   4,   relevant
formula's  for  $J/\psi$, DY and MB cross sections will be noted.
Analysis of  NA50  data  on  the  centrality  dependence  of  the
ratio's,  $J/\psi$/MB, DY/MB and $J/\psi$/DY will be presented in
section  5.  In  section  6  we  will  show  predicted   $J/\psi$
suppression  pattern  in Au+Au collisions at RHIC energy. Summary
and conclusions will be given in section 7.

\section{Model of $J/\psi$ production and absorption}

Qiu, Vary and Zhang \cite{qiu98} proposed a model to describe the
$J/\psi$     suppression    in    nucleon-nucleus/nucleus-nucleus
collisions.  For  the  sake  of  completeness,  we  will  briefly
describe  the  model.  Qiu,  Vary  and  Zhang  assumed  that  the
production  of  $J/\psi$  meson  is  a  two  step  process,   (i)
production  of  $c\bar  c$  pairs  with  relative momentum square
$q^2$, and (ii) formation of $J/\psi$ mesons from the  $c\bar{c}$
pairs.  Step  (i) can be accurately calculated in QCD. The second
step,  formation  of  $J/\psi$  mesons  from  initially   compact
$c\bar{c}$ pairs is non-perturbative. They used a parametric form
for  the  step (ii), formation of $J/\psi$ from $c\bar{c}$ pairs.
The $J/\psi$ cross section in $AB$ collisions, at center of  mass
energy $\sqrt{s}$ was then written as,

\begin{equation}\label{1} \sigma_{A+B \rightarrow J/\psi + X} (s)
= K \sum_{a,b} \int dq^2 \left( \frac{\hat \sigma_{ab \rightarrow
cc}}     {Q^2}     \right)    \int    dx_F    \phi_{a/A}(x_a,Q^2)
\phi_{b/B}(x_b,Q^2) \frac{x_a x_b}{x_a + x_b} \times  F_{c\bar{c}
\rightarrow J/\psi} (q^2), \end{equation}

\noindent  where  $\sum_{a,b}$  runs over all parton flavors, and
$Q^2 = q^2 +4 m_c^2$. The  $K$  factor  takes  into  account  the
higher  order corrections. The incoming parton momentum fractions
are fixed by kinematics and are $x_a
=(\sqrt{x^2_F+4Q^2/s}+x_F)/2$               and              $x_b
=(\sqrt{x^2_F+4Q^2/s}-x_F)/2$.  Quark  annihilation   and   gluon
fusion  are the major sub processes for $c\bar{c}$ production. In
the leading log, they are given by \cite{be94},

\begin{eqnarray}   \label{2}  \hat  \sigma_{q\bar{q}  \rightarrow
c\bar{c}}  (Q^2)  =  &&\frac{2}{9}  \frac{4  \pi  \alpha_s}{3Q^2}
(1+\frac{\gamma}{2}) \sqrt{1-\gamma},\\
 \hat  \sigma_{gg  \rightarrow  c\bar{c}}  (Q^2)  = && \frac{ \pi
\alpha_s}{3Q^2} [(1+\frac{\gamma}{2} + \frac{\gamma^2}{16})  \log
(\frac{1+\sqrt{1-\gamma}}{1-\sqrt{1-\gamma}})                   -
(\frac{7}{4}+\frac{31}{16}\gamma)\sqrt{1-\gamma}] \end{eqnarray}

\noindent  where  $\alpha_s$ is the QCD running coupling constant
and  $\gamma  =  4  m_c^2/Q^2$.  In  Eq.\ref{1}   $F_{c   \bar{c}
\rightarrow  J/\psi}(q^2)$  is  the transition probability that a
$c\bar{c}$ pair with relative momentum square $q^2$ evolve into a
physical $J/\psi$  meson.  In  a  nucleon-nucleus/nucleus-nucleus
collision,  the  produced  $c\bar{c}$ pairs interact with nuclear
medium before they exit. Observed anomalous  nuclear  enhancement
of  the  momentum imbalance in dijet production led Qiu, Vary and
Zhang \cite{qiu98} to argue that the interaction of a  $c\bar{c}$
pair  with  nuclear  environment  increases  the  square  of  the
relative momentum between the $c\bar{c}$ pair. As a result,  some
of  the $c\bar{c}$ pairs can gain enough relative square momentum
to  cross  the  threshold  to  become  an   open   charm   meson.
Consequently,  the  cross  sections  for  $J/\psi$ production are
reduced in comparison with nucleon-nucleon cross section. If  the
$J/\psi$  meson travel a distance $L$, the transition probability
$F_{c\bar{c} \rightarrow  J/\psi}(q^2)$  in  Eq.\ref{1}  will  be
changed to,

\begin{equation}  \label{3}  F_{c\bar{c} \rightarrow J/\psi}(q^2)
\rightarrow F_{c\bar{c} \rightarrow J/\psi} (q^2 +  \varepsilon^2
L), \end{equation}

\noindent with $\varepsilon^2$ being the relative square momentum
gain per unit length. Qiu, Vary and Zhang \cite{qiu98} considered
three different parametric forms (representing different physical
processes)  for  the  transition probability. All the three forms
could  describe  the  experimental  energy  dependence  of  total
$J/\psi$  cross  section  in  hadronic  collisions  \cite{qiu98}.
However, only the following form,

\begin{equation} \label{4} F_{c \bar{c} \rightarrow J/\psi} (q^2)
=  N_{J/\psi} \theta(q^2) \theta({4m^\prime}^2 - 4 m_c^2 -q^2)
(1   -   \frac{q^2}{{4m^\prime}^2   -   4   m_c^2  })^{\alpha_F},
\end{equation}

\noindent  could  describe  the  experimental  $J/\psi$ data as a
function of effective nuclear length \cite{qiu98}. In  Fig.1,  we
have shown the fit obtained to NA50 experimental data \cite{ab97}
on  the  $J/\psi$  total cross section as a function of effective
nuclear length. We have used CTEQ5M parton distribution functions
\cite{cteq5}.   The   parameter   values,    $KN_{J/\psi}$=0.458,
$\varepsilon^2$=0.225  $GeV^2/fm$ and $\alpha_F$=1 are very close
to the values obtained in Ref.\cite{qiu98}.

\begin{figure}[h]
\centerline{\psfig{figure=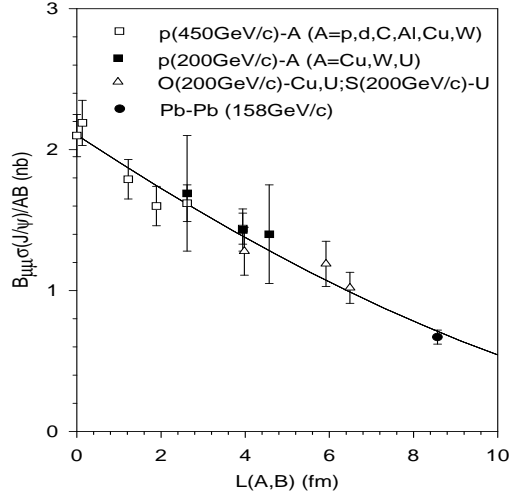,height=9cm,width=8cm}}
\vspace{-2cm}  \caption{  Total  $J/\psi$ cross sections with the
branching ratio to $\mu^+\mu^-$ in proton-nucleus, proton-nucleus
and nucleus-nucleus collisions, as a function  of  the  effective
nuclear length L(A,B).} \end{figure}

\section{Transverse Energy Distributions}

We  have  assumed  that the probability to obtain $E_T$ at impact
parameter ${\bf b}$ can be parameterized as,

\begin{equation}        \label{b1}        P(b,E_T)        \propto
exp(-(E_T-qN_p(b))^2/2q^2aN_P(b)) \end{equation}

The  mean  value of the transverse energy is $<E_T>(b)= qN_p(b)$,
where $q$ is the average transverse energy per  participant,  and
the  dispersion  is  $\sigma^2_{E_T}=aq^2N_p(b)$,  where $a$ is a
dimensionless parameter. The number of participants  $N_p(b)$  at
impact parameter $b$ is calculated as,

\begin{equation}   \label{b2}  N_p(b)=\int  d^2s  n_p({\bf  s,b})
\end{equation}

\noindent where the transverse density $n_p({\bf b,s})$ is,

\begin{equation}  \label{15}  n({\bf  b,s})  =  T_A({\bf s}) [1 -
e^{-\sigma_{NN}  T_B({\bf  b-s})}]  +   T_B({\bf   b-s})   [1   -
e^{-\sigma_{NN} T_A({\bf s})}], \end{equation}

Nuclear    thickness    function,   $T_{A,B}({\bf   s})=\int   dz
\rho_{A,B}({\bf s},z)$ contain all the  nuclear  information.  In
the  present  calculation,  we have used the following parametric
form for $\rho_A(r)$ \cite{bl00},

\begin{equation}                                        \label{6}
\rho_A(r)=\frac{\rho_0}{1+exp(\frac{r-r_0}{a})} \end{equation}

\noindent  with  a=0.53 fm, $r_0=1.1A^{1/3}$. The central density
is obtained from $\int \rho_A(r)d^3r=A$.

The  parameters  $q$  and  $a$  were  obtained  by  fitting  NA50
experimental transverse energy distribution.  NA50  collaboration
\cite{na50}  did not correct the $E_T$ spectra for the efficiency
of target identification algorithm, which is lower than unity for
$E_T$ lower than 60 GeV. To obtain the parameters $q$ and $a$  we
have  fitted the purely inclusive part of the $E_T$ spectra ($E_T
>$ 60 GeV). The fit is shown  in  Fig.2.  It  was  obtained  with
$q$=0.274 GeV and $a$=1.97.

\begin{figure}[h]
\centerline{\psfig{figure=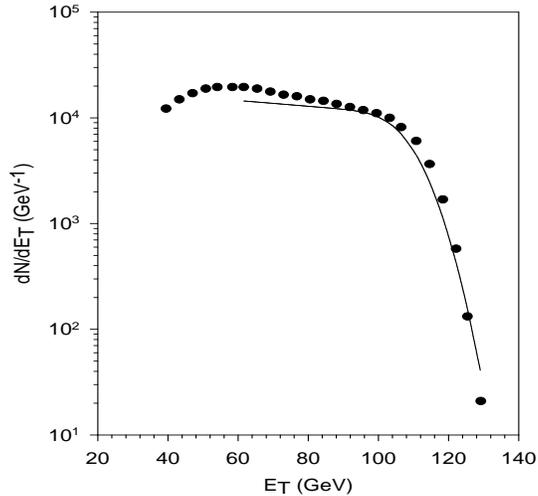,height=9cm,width=8cm}}
\vspace{-2cm}     \caption{Experimental     transverse    energy
distribution in Pb+Pb collisions  along  with  the  fit  obtained
assuming Gaussian $E_T-b$ correlation.} \end{figure}

\section{$E_T$  dependence of $J/\psi$, DY and MB events}

Here  we present in brief, the relevant formula's for calculating
minimum bias, Drell-Yan and $J/\psi$ cross sections.

Minimum  bias  cross  sections  are  easy  to  calculate.  It  is
essentially the inelastic cross-section. In the Glauber model, at
impact parameter ${\bf b}$ minimum  bias  cross  section  can  be
calculated as,

\begin{equation}  \label{12} d^3\sigma^{MB}/dE_T d^2b = (1 - exp(
-\sigma_{NN} T_{AB}(b)) P(b,E_T) \end{equation}

The  Drell-Yan  pairs  do not suffer any final state interactions
and the cross section at an impact parameter  ${\bf  b  }$  as  a
function of $E_T$ can be written as,

\begin{equation}    \label{13}    d^3\sigma^{DY}/dE_T    d^2b   =
\sigma_{NN}^{DY} \int d^2s  T_A({\bf  s})  T_B({\bf  s}-{\bf  b})
P(b,E_T), \end{equation}

\noindent where $\sigma_{NN}^{DY}$ is the Drell-Yan cross section
in $NN$ collisions.

While  Drell-Yan  pairs  do  not suffer interactions with nuclear
matter, the $J/\psi$ mesons do. In the present model  suppression
factor depend on the length traversed by the $c\bar{c}$ mesons in
nuclear medium. Consequently, we write the $J/\psi$ cross section
at an impact parameter ${\bf b}$ as,

\begin{equation}   \label{14}   d^3\sigma^{J/\psi}/dE_T   d^2b  =
\sigma_{NN}^{J/\psi}  \int  d^2s  T_A({\bf  s})  T_B({\bf   s-b})
S(L({\bf b,s})) P(b,E_T), \end{equation}

\noindent  where  $\sigma_{NN}^{J/\psi}$  is  the  $J/\psi$ cross
section  in  $NN$  collisions  and  $S(L({\bf  b,s}))$   is   the
suppression  factor  due to passage through a length L in nuclear
environment. The length $L({\bf b,s})$ that  the  $J/\psi$  meson
will traverse can be obtained as,

\begin{equation}  \label{16}  L({\bf  b,s})=n({\bf b,s})/2 \rho_0
\end{equation}

\noindent where $n(b,s)$ is the transverse density [Eq.\ref{15}].
Suppression  factor  $S(L({\bf  b,s})$  can  be  calculated using
Eq.\ref{1}, noting that a $c\bar{c}$ pair gains  relative  square
momentum  $\varepsilon^2$,  traversing  a  square  of length $L$.
Parametric  value  of  $\varepsilon^2$,  as  shown  before,   was
obtained  by fitting nucleon-nucleus and nucleus-nucleus $J/\psi$
cross section data containing  all  $E_T$.  However,  Eq.\ref{14}
corresponds  to  a particular $E_T$. Accordingly, square momentum
gain factor needs to be modified. We  modify  the  momentum  gain
factor  $\varepsilon^2$ to take into account the $E_T$ dependence
as,

\begin{equation}   \label{17}  \varepsilon^2(E_T)=\varepsilon^2_0
\frac{L(E_T)}{<L>} , \end{equation}

\noindent where $\varepsilon^2_0$ is the momentum gain factor for
all  $E_T$  (which  was  obtained  by fitting experimental data).
$L(E_T)$ is the  length  through  which  a  $J/\psi$  meson  with
transverse  energy  $E_T$ will travel. The length $L(E_T)$ can be
calculated as \cite{vo99},

\begin{equation}   \label{18}   L(E_T)=  \frac{  \int  d^2b  d^2s
T_A({\bf  s})  T_B({\bf  b-s})  [T_A({\bf   s})+T_B({\bf   b-s})]
P(b,E_T)  }  {2 \rho_0 \int d^2b d^2s T_A({\bf s}) T_B({\bf b-s})
P(b,E_T)}. \end{equation}

Fluctuations  of  transverse  energy  at a fixed impact parameter
plays an important role in the  explanation  of  the  NA50  data.
Above  100  GeV, i.e., approximately at the position of the knee,
the 2nd drop in the data is due to the fluctuations in $E_T$ . In
order to account for the fluctuations, following  Capella  et  al
\cite{ca00}, we calculate,

\begin{equation}\label{19}    F(E_T)    =   E_T   /E_T^{NF}(E_T),
\end{equation}

\noindent where,

\begin{equation}   \label{20}  E_T^{NF}(E_T)  =  \frac{\int  d^2b
E_T^{NF}(b) P(b,E_T)} {\int d^2b P(b,E_T)} \end{equation}

The   function   $F(E_T)$   is  unity  up  to  the  knee  of  the
distribution,  and  increases  thereafter,  precisely  where  the
fluctuations dominate. The replacement,

\begin{equation} \label{21} L({\bf b,s}) \rightarrow L({\bf b,s})
F(E_T), \end{equation}

\noindent  then  properly  accounts  for  the fluctuations in the
$E_T$ distributions.

\section{Comparison with experiments}

Capella  et  al  \cite{ca01}  observed  that  in a $J/\psi$ event
sample and also in a DY event sample, $E_T \sim$ 3 GeV  is  taken
by  the  trigger  and thus the transverse energy deposited in the
calorimeter by the  other  hadrons  is  slightly  less  than  the
corresponding  one in a minimum bias event sample. The $E_T$ loss
in a $J/\psi$ event, though very small, can have larger effect in
the  ratio  of  $J/\psi$  over  MB  at  the  tail  of  the  $E_T$
distribution.  However,  in  the  ratio  of  $J/\psi$ over DY the
effect  of  $E_T$ loss cancels out. Exact magnitude of $E_T$ loss
in a $J/\psi$ event is  difficult  to  estimate.  Capella  et  al
\cite{ca01} single out the NN collision in which $J/\psi$ (or DY)
is  produced and assumed that in that collision, no other hadrons
are produced. We have implemented this in the present calculation
by replacing the number of participants $N_p(b)$ in a $J/\psi$ or
DY event by,

\begin{equation} N_p(b) \rightarrow N_p(b) - 2, \end{equation}

\noindent minimum number of participants being two.

\begin{figure}[h]
\centerline{\psfig{figure=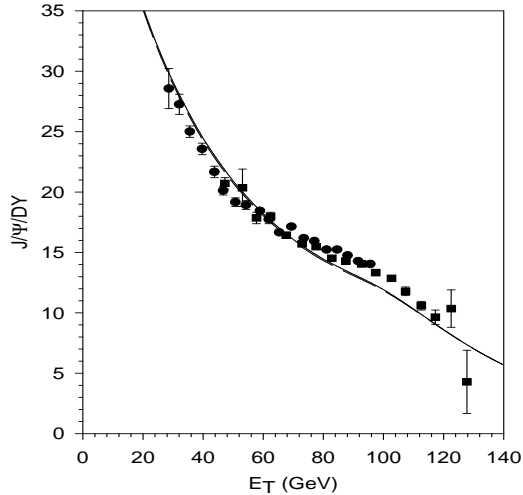,height=9cm,width=8cm}}
\vspace{-2cm} \caption{Filled and square circles are the $J/\psi$
to   Drell-Yan  ratio  in  Pb+Pb  collisions,  obtained  by  NA50
collaboration in 1996 and 1998 respectively. The  solid  line  is
the   model  prediction  with  or  without  $E_T$  loss  effect.}
\end{figure}

In   Fig.3,  we  have  compared  the  model  prediction  for  the
transverse energy dependence of $J/\psi$  over  Drell-Yan  ratio,
with   the   NA50   experimental   data.   We  have  assumed  the
normalization      factor      $\sigma^{J/\psi}/\sigma^{DY}$=53.5
\cite{bl00}.  The  solid  line  is the model prediction, obtained
without $E_T$ loss consideration. Identical  result  is  obtained
when  $E_T$  loss  effect  is  taken  into  consideration. In the
$J/\psi$ over DY ratio,  the  effect  get  cancelled  to  a  high
accuracy.  The model gives good description to the NA50 data. The
2nd drop in the ratio at 100 GeV  is  reproduced.  We  note  that
without  $E_T$  fluctuations,  that  drop  is not reproduced. The
result clearly shows  that  it  is  not  essential  to  assume  a
deconfined  scenario  to  explain  the  NA50  data  on centrality
dependence of $J/\psi$ to  Drell-Yan  ratio.  Nuclear  absorption
alone is capable of explaining the data.

\begin{figure}[h]
\centerline{\psfig{figure=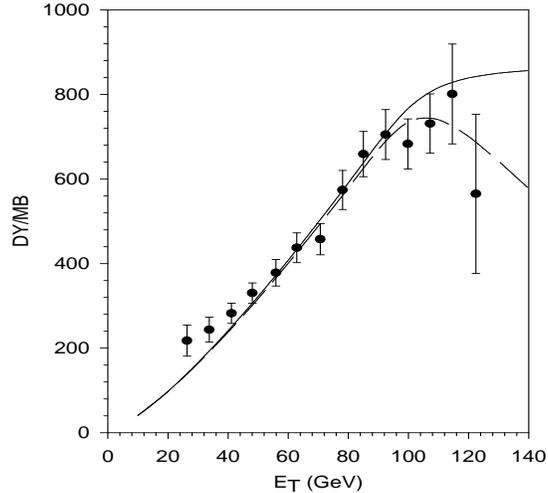,height=9cm,width=8cm}}
\vspace{-2cm}  \caption{NA50 data on the centrality dependence of
the ratio DY/MB is compared with present model. The solid line is
the model calculation without $E_T$ loss effect. The dashed  line
incorporate the $E_T$ loss effect.} \end{figure}

In  Fig.4,  NA50 data \cite{na50_99} on the centrality dependence
of the Drell-Yan over minimum bias ratio is shown. The data  were
taken  in  1996 using a thick target. NA50 collaboration does not
consider  the  data  reliable  beyond  the  knee  of  the   $E_T$
distribution,  due  to  possible contamination of rescattering in
the thick target. The solid line is a fit obtained  to  the  data
without  $E_T$  loss  effect. The ratio saturates beyond 100 GeV.
When  effect  of  $E_T$  loss  in  a  DY  event  is  taken   into
consideration  (the  dashed line) the ratio drops beyond 100 GeV.
The effect is significant. At large $E_T$,  the  ratio  drops  by
more  than  60\%  when  $E_T$  loss  effect  is included. Indeed,
experimental data do show such a drop, but as  told  earlier  the
data  are  not reliable beyond 100 GeV. The prediction can easily
be verified in future experiments.

NA50   collaboration   reported   the   preliminary  analysis  of
centrality dependence of $J/\psi$ over MB  ratio  \cite{na50_01}.
In Fig.5, we have shown the preliminary data. The ratio increases
up  to  the  knee  of  the  $E_T$  distribution,  there  after it
decreases. There is an  indication  of  change  in  slope  around
$E_T$=40  GeV.  In  Fig.5,  dashed  and  solid  lines  are  model
calculations, obtained with and without $E_T$  loss  effect.  The
model  predictions  agree  with  experimental data. The change of
slope around $E_T$=40 GeV is also reproduced.
 As  with DY/MB ratio, the two predictions differ only beyond 100
GeV, where $E_T$ fluctuations dominate.  With  $E_T$  loss  taken
into  account,  the  ratio drops faster, fitting the data better.
Here also, the effect is  significant,  60\%  or  more  at  large
$E_T$.

NA50  collaboration  analyzed  the  $J/\psi$ data in two distinct
manner. One is the standard  analysis,  in  which  the  ratio  of
$J/\psi$ over Drell-Yan (DY) was measured. In the other, which is
called  minimum  bias  (MB)  analysis,  the Drell-Yan events were
replaced by the minimum bias events. The ratio of  $J/\psi$  over
minimum  bias  cross  section  was  then  divided by the ratio of
theoretical DY over minimum bias cross section to obtain

\begin{equation} {\left( \frac{J/\psi} {DY} \right ) }_{MB}
 =
\left    (    \frac{J/\psi}{MB}    \right    )_{EXP}    \left   (
\frac{MB}{DY}\right )_{TH} \end{equation}

The  standard  analysis  data  cover  only  up to the knee of the
$E_T$-distribution. The behavior of the $J/\psi$  over  DY  ratio
beyond  the  knee  of the $E_T$-distribution is obtained entirely
from the minimum bias analysis. Without $E_T$ loss effect,  DY/MB
remain  practically  constant  beyond  100 GeV, and therefore the
behavior of $J/\psi$ over DY and that  of  $J/\psi$  over  MB  is
practically  the  same.  However,  present  calculations  clearly
indicate that, due to $E_T$ loss effect, DY/MB  does  not  remain
constant  but  changes  quite  significantly in this $E_T$ range.
Then behavior of $J/\psi$ over MB and that of  $J/\psi$  over  DY
are  not  the same. True ratio will be flatter than obtained from
the minimum bias analysis. As suggested by Capella et al,  it  is
better not to superimpose two analysis in a single figure. Models
can be tested better by fitting them separately.

\begin{figure}[h]
\centerline{\psfig{figure=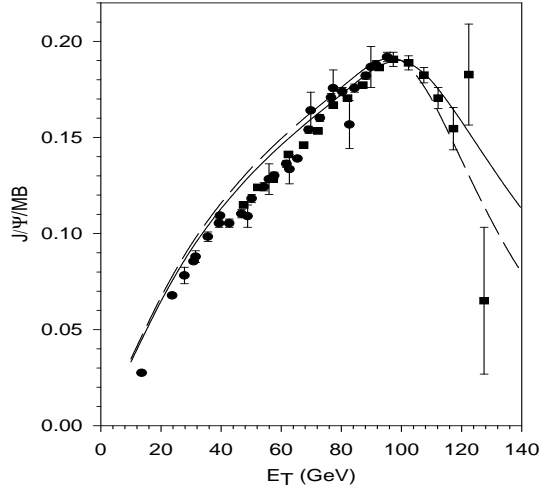,height=9cm,width=8cm}}
\vspace{-2cm}  \caption{NA50 (preliminary) data on the centrality
dependence of the ratio  $J/\psi$/MB  is  compared  with  present
model. The solid line is the model calculation without $E_T$ loss
effect.  The  dashed  line  incorporate  the  $E_T$ loss effect.}
\end{figure}

\section{Prediction for RHIC energy}

Present  model  can  be  used to predict centrality dependence of
$J/\psi$ over DY in Au+Au collisions at RHIC energy. Experimental
$E_T$ distribution at RHIC energy is not available,  but  we  can
make an educated guess for it.

\begin{figure}[h]
\centerline{\psfig{figure=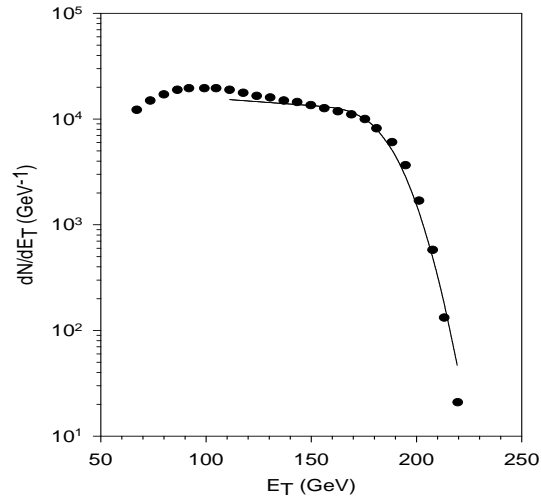,height=9cm,width=8cm}}
\vspace{-2cm}  \caption{Simulated  $E_T$  distribution  for Au+Au
collisions  at  RHIC  energy  is  fitted  with  Gaussian  $E_T-b$
correlation.} \end{figure}

 Recent  PHOBOS  experiment \cite{phobos} showed that for central
collisions, total multiplicity is larger by 70\% at RHIC than  at
SPS.   We   assume   that   $E_T$  is  correspondingly  increased
\cite{bl01}. Accordingly, we rescale the $E_T$  distribution  for
Pb+Pb  collisions  and  assume that it represent the experimental
$E_T$ distribution for  Au+Au  collisions  at  RHIC  (small  mass
difference  between  Au and Pb is neglected). At RHIC energy, the
so-called hard component, which is proportional to the number  of
binary  collision, appears. Model dependent calculations indicate
that the hard component grows from 22\% to  37\%  as  the  energy
changes from $\sqrt{s}$=56 GeV to 130 GeV \cite{kh01}. However we
choose  to  ignore  the hard component in $E_T$-distribution. The
multiplicity distribution obtained by the  PHOBOS  collaboration,
in  the  rapidity  range  $3<\mid \eta \mid <4.5$ could be fitted
well with or without this hard component. Indeed, it appears that
the data are  fitted  without  the  hard  component  \cite{kh01}.
Global   distribution  e.g.  multiplicity  or  transverse  energy
distributions are not sensitive to the hard component.

In  fig.6,  filled  circles  represent  the  "experimental" $E_T$
distribution for Au+Au collisions at RHIC,  obtained  by  scaling
the $E_T$ distribution in Pb+Pb collisions at SPS. The solid line
is  a  fit to the "experimental" $E_T$-distribution obtained with
$q$=.46 GeV and $a$=1.97. Nucleon-nucleon inelastic cross section
($\sigma_{NN}$) was assumed to be 41 mb at RHIC, instead of 32 mb
at SPS \cite{bl01}.

\begin{figure}[h]
\centerline{\psfig{figure=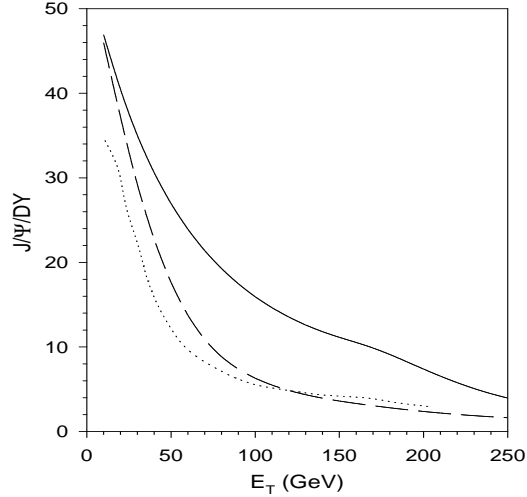,height=9cm,width=8cm}}
\vspace{-2cm}   \caption{Predicted   centrality   dependence   of
$J/\psi$/DY ratio at RHIC energy for Au+Au collisions. The dashed
and solid lines are present model predictions  with  and  without
hard  scattering  component in the transverse density. The dotted
line is obtained in a model where all the $J/\psi$'s melts  above
a threshold density.} \end{figure}

While the $E_T$ distribution at RHIC energy may not depend on the
so-called  hard  component  which  is  proportional  to number of
binary collisions, $J/\psi$ suppression will strongly  depend  on
the  hard  component,  as it effectively increases the density of
the  nuclear  medium.  For  $f$  fraction  of  hard   scattering,
transverse  density  $n({\bf b,s})$ in Eq.\ref{15} is modified to
\cite{ch02},

\begin{equation} \label{22} n_{mod} ({\bf b,s}) \rightarrow (1-f)
n({\bf b,s}) + f n^{hard}({\bf b,s}) , \end{equation}

\noindent  with  $n^{hard}({\bf  b,s})=\sigma_{NN}  T_A({\bf  s})
T_B({\bf b-s})$.  With  hard  component,  transverse  density  is
increased,  as  a  result,  suppression will be increased at RHIC
energy.  In  Fig.7,  solid  line  is  the  predicted   centrality
dependence  of  the $J/\psi$/DY ratio without any hard scattering
component in  the  transverse  density.  Suppression  pattern  is
similar  to  that in Pb+Pb collisions at SPS energy. The 2nd drop
now shifts at higher $E_T$. With 37\% hard  scattering  component
in the transverse density (the dashed line) suppression increases
and  the  effect  of  $E_T$ fluctuations, i.e. the second drop is
washed out. In Fig.7, we have shown the prediction (dotted  line)
obtained  in  a model where all the $J/\psi$'s melts down above a
threshold density \cite{bl01}. The prediction also included  37\%
hard  scattering.  It  agrees  closely  with  the  present  model
prediction (dashed line). Assumption of $J/\psi$'s melting  above
a  threshold  density  implicitly  assume QGP like environment is
produced in the collision. Very similar prediction  for  $J/\psi$
suppression  in a confined scenario and in a deconfined scenario,
raises the possibility that it may not be possible to confirm the
deconfinement phase transition, even at RHIC energy, from  the
$J/\psi$  data.  Recently  several  authors have proposed that at
RHIC  energy,  in  a  deconfined   scenario,   recombination   of
$c\bar{c}$  pairs  will lead to enhancement of $J/\psi$'s, rather
than its suppression  \cite{recom}.  Inclusion  of  recombination
effects  may  mask the large suppression obtained in a deconfined
scenario. However, nuclear suppression  as  calculated  presently
will remain unaltered. It may then be possible to distinguish the
deconfinement phase transition from the $J/\psi$ data.

We  have also calculated the centrality dependence of $J/\psi/MB$
and of $DY/MB$ ratio's at RHIC energy. Centrality  dependence  of
$DY/MB$  is  very similar to that obtained in Pb+Pb collisions at
SPS energy and is not shown. Centrality  dependence  of  $J/\psi$
over  MB  ratio  is  shown in Fig.8. The upper panel was obtained
without any hard scattering component in the transverse  density,
while in the lower panel, hard scattering (37\%) was included. As
before,  effect of $E_T$ loss is observed only beyond the knee of
the $E_T$ distribution. It is interesting to note the  difference
in the shape of the ratio. Without hard scattering component, the
ratio   increases  with  $E_T$  up  to  the  knee  of  the  $E_T$
distribution, then decreases, a dependence similar to Pb+Pb  data
at SPS energy. With hard scattering included, the ratio increases
at  low  $E_T$,  then remains more or less same up to the knee of
the $E_T$ distribution. Beyond the knee, it decreases again. Very
distinct shape of the two distributions can be used to  determine
experimentally the fraction of hard processes in the collision.

\begin{figure}[h]
\centerline{\psfig{figure=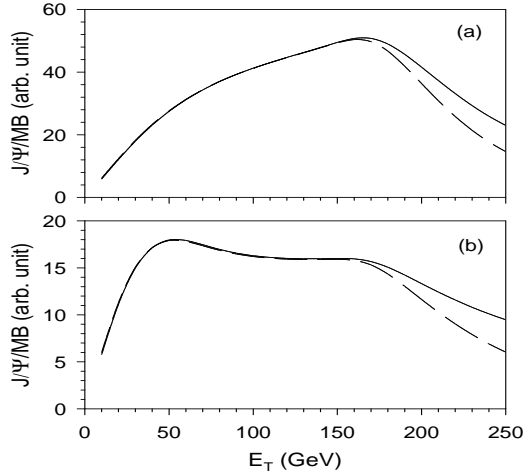,height=9cm,width=8cm}}
\vspace{-2cm}  \caption{Centrality  dependence  of $J/\psi$/MB at
RHIC energy. Upper panel was  obtained  without  hard  scattering
while  37\%  hard  scattering is included in the lower panel. The
solid and dashed lines are obtained without and with  $E_T$  loss
effect.} \end{figure}

\section{Summary and Conclusions}

We  have  studied  the NA50 data on $J/\psi$ suppression in Pb+Pb
collisions at SPS energy. The $J/\psi$ production is  assumed  to
be  a two step process, (i) production of $c\bar{c}$ pairs, which
is accurately calculable in pQCD and (ii) formation  of  $J/\psi$
from  the  $c\bar{c}$ pair, essentially non-perturbative, but can
be conveniently parameterized. In  the  model,  $c\bar{c}$  pairs
gain  relative  square  momentum as is travel through the nuclear
environment, and  some  of  the  pairs  can  gain  enough  square
momentum  to  cross  the threshold to become an open charm meson.
The parameters of the model were fixed  by  fitting  experimental
total  $J/\psi$ cross section in pp,pA and AA collisions. Without
any free parameter,  the  model  could  very  well  describe  the
centrality  dependence  of  the $J/\psi$ over DY ratio. The model
also reproduces the centrality dependence of the $J/\psi/MB$  and
$DY/MB$ ratio. We have also studied the effect of $E_T$ loss in a
$J/\psi$ or DY event. Those events on the average produces $\sim$
3  GeV  less $E_T$ than a minimum bias event. As shown by Capella
et al \cite{ca01}, the effect of $E_T$ loss get cancelled in  the
ratio  $J/\psi/DY$.  But  it  affect the $J/\psi/MB$ ratio. $E_T$
loss also  affect  the  $DY/MB$  ratio.  Ratio  does  not  remain
constant beyond the knee of the $E_T$ distribution, but drops.

We  have also used the model to predict the centrality dependence
of the $J/\psi$ over DY ratio at RHIC  energy.  At  RHIC  energy,
hard  scattering component may be effective. $J/\psi$ suppression
depends strongly on the hard scattering  component.  Without  any
hard   scattering,   $J/\psi$   suppression  closely  follow  the
suppression pattern at SPS energy.  The  2nd  drop  occurring  at
higher $E_T$. However, suppression increases when hard scattering
is included and washes out the 2nd drop in the ratio. Suppression
obtained  in  the  nuclear  environment  closely agree with model
prediction  where  QGP  formation  is  implicitly  assumed.  Hard
scattering also affect the $J/\psi$/MB ratio very strongly. While
without  any hard scattering, $J/\psi/MB$ increases up to knee of
the $E_T$ distribution, with hard scattering included, the  ratio
remains  more  or  less  same  over  a  large range of $E_T$. The
distinct  change  in  shape  raises  the  possibility  of   using
$J/\psi/MB$  ratio  to find the fraction of hard processes in the
collision.

To  conclude, it is not essential to assume a deconfined scenario
to  explain  the  anomalous  suppression  of  $J/\psi$  in  Pb+Pb
collisions  at  CERN  SPS. The data are explained in terms normal
nuclear suppression, also present in pa/AA collisions.  Predicted
suppression  pattern at RHIC energy indicate, that even there, it
may not be possible to identify  deconfinement  phase  transition
from the $J/\psi$ data.

The  author  would  like  to  thank  Prof.  A.  Capella  for  his
suggestions and comments.

 \end{document}